\documentclass[conference]{IEEEtran}
\IEEEoverridecommandlockouts
\usepackage{cite}
\usepackage{amsmath,amssymb,amsfonts}
\usepackage{braket}
\usepackage{algorithm} 
\usepackage{algpseudocode} 
\usepackage{graphicx}
\usepackage{tabularx}
\usepackage{float}
\usepackage{textcomp}
\usepackage{xcolor}
\usepackage{url}
\usepackage{hyperref}
\hypersetup{colorlinks=true, citecolor=blue,linkcolor=blue,filecolor=blue,urlcolor=blue}

\def\BibTeX{{\rm B\kern-.05em{\sc i\kern-.025em b}\kern-.08em
    T\kern-.1667em\lower.7ex\hbox{E}\kern-.125emX}}

\newcommand{\Lts}{\mathcal{L}}
\newcommand{\abort}{\bot}
\newcommand{\Cons}[2]{\mathsf{Cons}_{#1}(#2)}
\newcommand{\cmd}[1]{\mathsf{cmd}(#1)}
\newcommand{\contrad}[2]{\big(\cmd{#1}\neq \cmd{#2}\big)}

\begin{document}

\title{\textit{In Silico} Benchmarking of Detectable Byzantine Agreement in Noisy Quantum Networks}

\author{\IEEEauthorblockN{Mayank Bhatia, Shaan Doshi}
\IEEEauthorblockA{\textit{Siebel School/NCSA} \\
\textit{UIUC}\\
Urbana IL, USA \\
\{mayankb3,shaand3\}@illinois.edu}
\and
\IEEEauthorblockN{Daniel Winton, Brian Doolittle}
\IEEEauthorblockA{\textit{Aliro Technologies} \\
Brighton, MA, USA \\
\{daniel,bdoolittle\}@aliroquantum.com}
\and
\IEEEauthorblockN{Bruno Abreu}
\IEEEauthorblockA{\textit{PSC} \\
Pittsburgh PA, USA \\
babreu@psc.edu}
\and
\IEEEauthorblockN{Santiago Núñez-Corrales}
\IEEEauthorblockA{\textit{NCSA/IQUIST} \\
\textit{UIUC}\\
Urbana IL, USA \\
nunezco2@illinois.edu}
}

\maketitle

\begin{abstract}
Quantum communication resources offer significant advantages for fault-tolerant distributed protocols, particularly in Byzantine Agreement (BA), where reliability against adversarial interference is essential. Quantum Detectable Byzantine Agreement (QDBA) enables consensus protocols that surpass classical limitations by leveraging entangled quantum states. In this work, we focus on the practical realization of QDBA using Einstein-Podolsky-Rosen (EPR) pairs, the simplest maximally entangled quantum resources, making the protocol experimentally accessible across current quantum hardware platforms. We present a comprehensive computational study of the EPRQDBA protocol under realistic quantum network conditions, utilizing the Aliro Quantum Network Simulator to evaluate the performance and robustness of the protocol. Our simulations systematically explore the protocol's parameter space—including variations in network size, traitorous node count, the amount of entanglement consumed in the protocol, and physically motivated noise models tailored specifically for superconducting and photonic qubit technologies. Through extensive numerical experiments, we provide insights into how these physically realistic parameters impact protocol performance, establishing critical thresholds and optimal operational regimes for experimental implementations. This work bridges theoretical advances in quantum consensus protocols with practical network implementations, offering a concrete reference for experimentalists. Our findings serve as a guideline for evaluating and optimizing QDBA implementations in realistic, noisy environments.
\end{abstract}



\section{Introduction}







Distributed consensus is a fundamental and well-studied topic across multiple research areas, including robotics \cite{fengying2024decentralized}, optimization \cite{tsianos2012consensus}, distributed systems \cite{nguyen2002consensus}, and networking \cite{liu2014using}. The focus of the topic is coordinating multiple agents for cooperative control of a system despite some of the agents being faulty. A particular area of interest within distributed consensus is the Byzantine Agreement (BA) problem \cite{pease1980_reaching_agrmnt_with_faults,Lamport1982, dolev1983_authenticated_ba,castro1999practical_cba,castro2002_practical_cba,Aublin2013_redundant_ba,miller2016_honeybadger_ba,yin2019_hotstuff_ba, Lu2020_mvba}, which aims to achieve unity across all non-faulty processes even when faulty processes can be adversarial and follow a coordinated malevolent plan. Solutions to the BA problem, often referred to as Byzantine fault tolerant protocols, have been applied in algorithms for distributed sensing \cite{Brooks1996_ba_distributed_sensing,cluster-wireless-sensor-net}, clock synchronization \cite{clock-sync}, and blockchain technologies \cite{miller2016_honeybadger_ba,yin2019_hotstuff_ba,tholoniat2022_ba_blockchain}. Classically, BA has long been known to be impossible if the adversary controls $T > \frac{n}{3}$ agents and we assume only pairwise authenticated channels and no cryptographic identity management.

A relaxation of BA called Detectable Byzantine Agreement (DBA) is satisfied when a loyal agent either agrees with other loyal nodes or aborts the protocol if foul play is detected. DBA is achieved under the following conditions: (1) All non-faulty agents agree on the same value. (2) An agent can abort from the protocol, which does not affect the success of the protocol. DBA can be trivially achieved by all non-faulty players aborting \cite{Fitzi_2001_first_QBA}. So, we are interested in DBA protocols which abort minimally and (almost) never fail.

Quantum communication resources, such as shared entanglement or point-to-point quantum communication channels, are known to provide advantages in achieving DBA \cite{Fitzi_2001_first_QBA,Cabello2002_n-level_singlet_applications,Cabello2003_liar_detection_state,Cabello2003_liar_detection_problem,Iblisdir2004_ba_with_qkd_setup,EPRQDBA}.
 Several quantum detectable byzantine agreement (QDBA) protocols highlight that using quantum entanglement allows for  $T > \frac{n}{3}$ malicious agents to be tolerated using only pairwise authenticated channels and also does not rely on an assumption of cryptographic identity management, therefore supporting both an arbitrary number of faulty nodes and adversaries with unbounded computational resources. QDBA protocols have also been demonstrated experimentally in optical setups \cite{Gaertner2008_expt_qdba,Weng2023_quantum_byzantine_agreement,experimental_3_photonic} and quantum processors \cite{prest2023_qdba_on_quantum_processor}.

This work focuses on the QDBA protocol from Andronikos and Sirokofskich \cite{EPRQDBA}, which uses only pairwise entanglement to establish DBA among an arbitrary number of parties. We are interested in this protocol for its practicality because bipartite entanglement is the most basic and easiest form of entanglement to generate in practice systems. Other proposed QDBA protocols are also interesting, but require more complicated entanglement structures. The EPR-based QDBA protocol (EPRQDBA) is as simple as possible in terms of the quantum network which needs to be implemented. Furthermore, in noiseless conditions with appropriate internal tolerances, it almost never fails to achieve DBA. We are interested in how this protocol performs in common noisy quantum network setups such as superconducting and photonic systems. Particularly, we are interested in exploring the expected effect of different noises on the probability that a non-faulty agent aborts when it should not, as well as the probability that a non-faulty agent does not abort when it should. We also care to look into if adversaries can reliably coordinate successful attacks in noisy environments.

We use simulations to support experimentalists interested in implementing QDBA. Our results offer theoretical expectations for the performance of the protocol in different physical setups. In particular, we explore the effect of common logical noise models (e.g., amplitude damping), noise models pertaining to photonic setups, and noise models pertaining to superconducting circuits. Our goal is quantum networks hardware-protocol co-design. In particular, we make the following contributions: 

\begin{itemize}
    \item Exploration of parameter space using simulation to understand sensitivity of the model and construct useful a design of experiments connected to actual implementations
    \item Providing a computational rendition that focuses on the effect of the composition of the quantum network vis-a-vis the impact of expected noise sources on QDBA protocol security
\end{itemize}

\section{Quantum Byzantine Agreement}

\subsection{Significance of Byzantine Agreement}

Byzantine Agreement (BA) \cite{pease1980_reaching_agrmnt_with_faults,Lamport1982} is a fundamental primitive in fault-tolerant distributed computing that enables a group of participants to reliably reach consensus, even when some participants behave maliciously (Byzantine actors). The ability to achieve agreement under adversarial conditions is essential for ensuring consistency, reliability, and integrity across decentralized systems. This necessity becomes particularly evident in real-world applications involving untrusted or compromised communication channels.


Detectable Byzantine Agreement (DBA) extends traditional BA by introducing the capability to detect adversarial interference explicitly. In DBA protocols, participants can forfeit decisions upon detecting inconsistent or malicious behavior rather than being forced into an incorrect consensus. This capability significantly enhances accountability and system robustness, offering a meaningful safeguard beyond traditional consensus guarantees.

\subsection{QDBA review}

Classical Byzantine Agreement protocols typically rely on cryptographic hardness assumptions to achieve success, with exceptions such as common coin protocols \cite{feldman1988optimal}. However, they fail in the presence of an enemy with unbounded computational resources. Quantum Byzantine Agreement (QBA) addresses this vulnerability by providing information-theoretic security without depending on computational assumptions susceptible to quantum attacks. QBA protocols leverage quantum entanglement to achieve superior fault tolerance and security, making them inherently robust against quantum adversaries. Traditional quantum BA protocols often rely on complex multipartite entanglement (such as $n$-partite GHZ states) which are hard to achieve experimentally. Recent protocols leverage simpler entangled resources, specifically Einstein-Podolsky-Rosen (EPR) pairs, thereby significantly simplifying experimental implementation. Such simplification makes the protocol more accessible for experimentalists, promoting practical testing and validation on emerging quantum hardware platforms.

\subsection{Summary of a EPR-based QDBA protocol}


The QDBA protocol consists of a single preselected commander and $N-1$ lieutenants. There are three overarching steps: (1) entanglement distribution, (2) the commander sends orders and a proof vector based on the qubits it received, (3) lieutenants discuss classically to share confidence in the correct decision or suscpicion to abort. In the quantum-classical hybrid network, any player can exchange classical messages with all others, while receiving both quantum and classical information for a single (not necessarily trusted) distributor.

\begin{figure}[h]  
    \centering
    \includegraphics[width=0.7\linewidth]{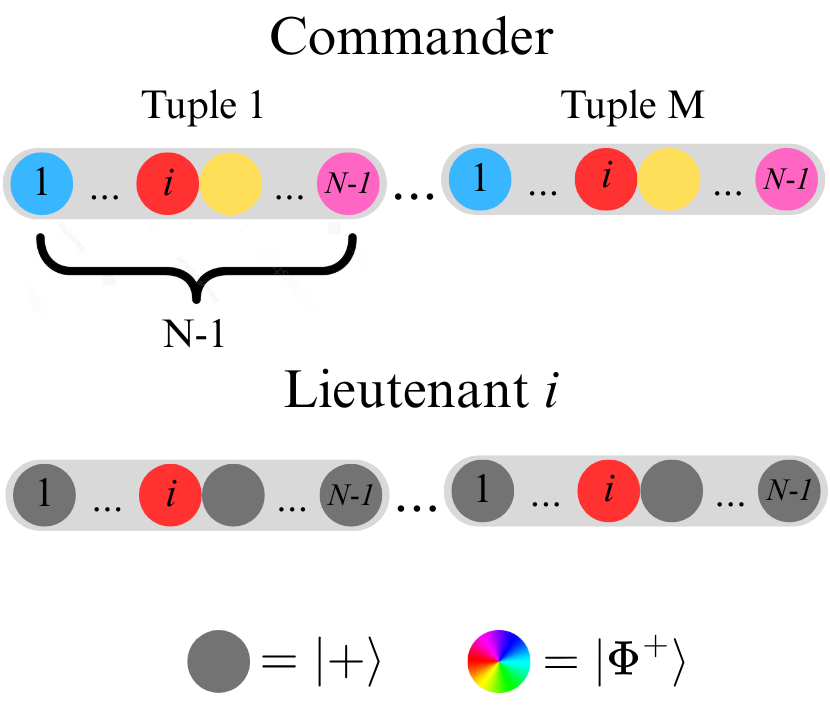}
    \caption{Illustration of EPRQDBA Entanglement Distribution. Same-color qubits represent anticorrelated qubits from the same EPR pair.}
    \label{fig:entanglement-distribution}
\end{figure}

The entanglement distribution step is illustrated in figure \ref{fig:entanglement-distribution}.

A central trusted distributor generates $M(N-1)$ maximally entangled and anticorrelating pairs (e.g., the state $\ket{\Psi^+} = \frac{1}{\sqrt{2}}(\ket{01} + \ket{10})$), where $M$ is the protocol robustness parameter. The trusted distributor uses quantum channels to send the commander the first half of every entangled pair, and lieutenant $i = 0, 1, \dots, N-2$  is sent the second half of the entangled pair at index $p = 0, 1, \dots, M(N-1)-1$ satisfying $p \bmod N-1 \equiv i$. These anticorrelated qubits fall in indices $i + (N-1)k$ with $k = 0, 1, \dots, M-1$ for Lieutenant $i$, with $\ket{+}$ states being sent in all other indices as filler. These qubits can be sent one at a time to minimize quantum channels and memories.  Each succession of $N-1$ qubits received by a player is part of the same tuple, and there are $M$ tuples in total. In every tuple, the commander will now have one anticorrelated qubit with each lieutenant.

\begin{figure}[h]
    \centering
    \includegraphics[width=0.9\linewidth]{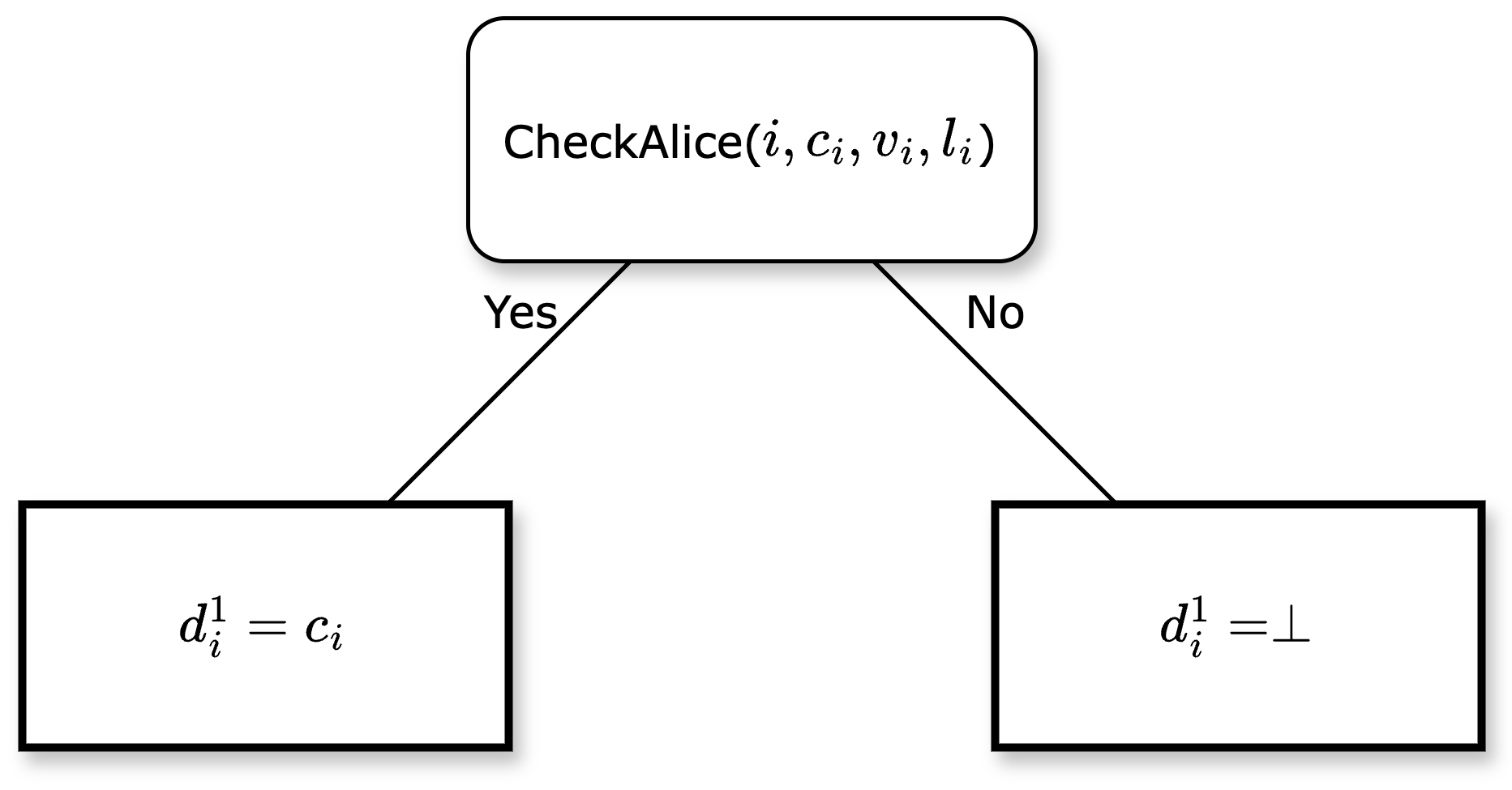}
    \caption{Round 1. The CheckAlice function attempts to verify if the received command vector $v_i$ is consistent with the received order $c_i$ for lieutenant $i$ with qubit measurements $l_i$. This is accomplished by expecting a reasonable number of unmasked tuples ($\approx \frac{1}{2}$) and anticorrelation in each tuple.}
    \label{fig:tree-1}
\end{figure}

\begin{figure}[h]
    \centering
    \includegraphics[width=0.9\linewidth]{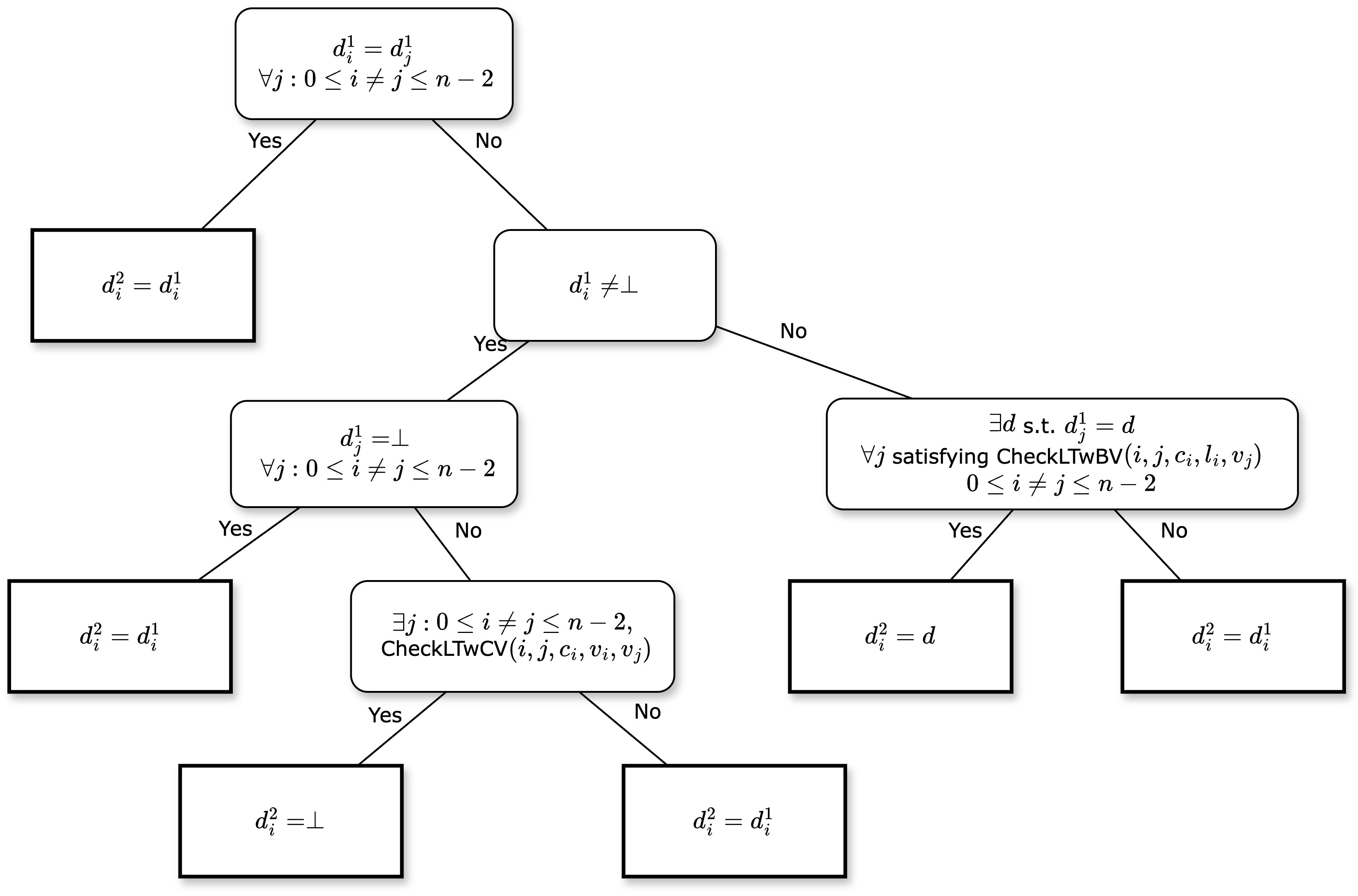}
    \caption{Round 2. The CheckLTwCV function attempts to determine if another lieutenant is loyal or Byzantine based on their supplied claim and proofs. It checks another lieutenant's command vector against the loyal lieutenant's bit vector to verify the expected correlations and anticorrelations. The CheckLTvBV is analogous, checking the other lieutenant's supplied bit vector with the loyal lieutenant's bit vector.}
    \label{fig:tree-2}
\end{figure}

\begin{figure}[h]
    \centering
    \includegraphics[width=0.9\linewidth]{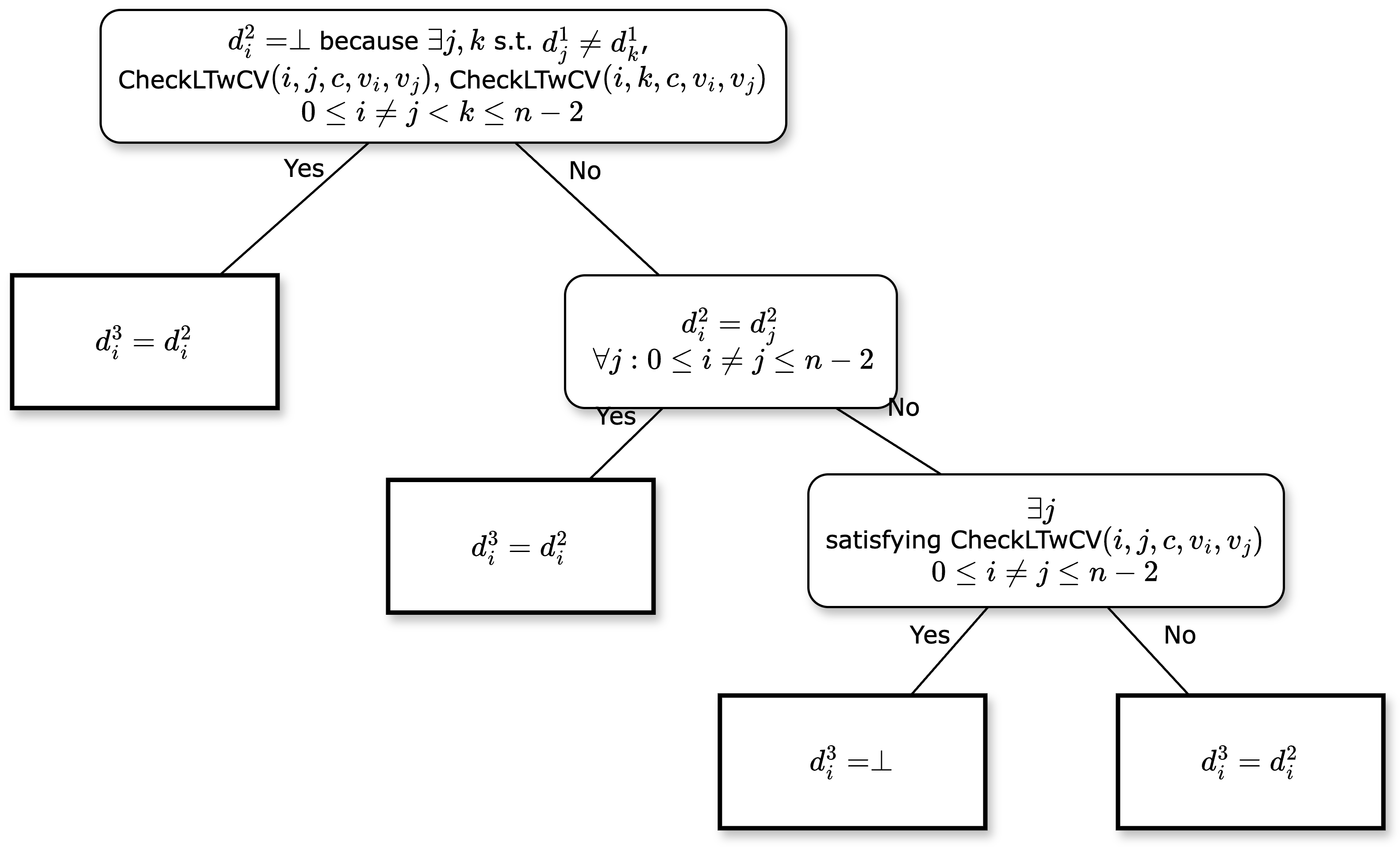}
    \caption{Round 3. Perform final santiy checks deciding whether or not to abort with the CheckLTwCV function. Else, cling onto the decision from round 2.}
    \label{fig:tree-3}
\end{figure}

Upon receiving the expected number of qubits ($M(N-1)$), the commander will issue a binary order to each lieutenant. This proof of this order is encoded in a masked version of the commander's measurement results called a command vector. Command vectors are unique for each lieutenant, even with the same order, and maintain the anticorrelated properties of the EPR pairs. The commander only reveals the a tuple to a lieutenant when the corresponding anti correlated qubit measurements match those in the command order. After the lieutenants receive their command vectors, they begin an all-to-all classical discourse to each decide what order to follow. The classical communication aspect of the protocol is described in Algorithm~\ref{alg:lieutenant} and the decision trees in Figs.~\ref{fig:tree-1}–\ref{fig:tree-3}. We write $\cmd{v}\in\{0,1\}$ for the order decoded from a command vector $v$. For brevity, the predicate $\Cons{i}{\cdot}$ always takes a command vector as its argument and checks it against lieutenant $i$’s available evidence.

\begin{algorithm}[h]
\caption{Lieutenant Protocol (for lieutenant $i\in\Lts$)}
\label{alg:lieutenant}
\begin{algorithmic}[1]
    \Procedure{LieutenantProtocol}{$i$}
        \State \textbf{(Setup)} receive $M(N-1)$ qubits and measure to obtain record $l_i$
        \State receive commander's order $c\in\{0,1\}$ and vector $v$
        \vspace{.25em}
        \Statex \textbf{Round 1}
        \State $d_i^1 \gets \begin{cases}
            c & \text{if } \Cons{i}{v} \\
            \abort & \text{otherwise}
        \end{cases}$ 
        \State send $(d_i^1, v)$ to all $j\in \Lts\setminus\{i\}$
        \State wait to receive $R_1^i=\{(d_j^1, v_j)\}_{j\in \Lts\setminus\{i\}}$
        \vspace{.25em}
        \Statex \textbf{Round 2}
        \If{$d_i^1 = c$ \textbf{ and } $\exists j\;[\,d_j^1=\overline{c}\ \land\ \Cons{i}{v_j}\,]$}
            \State $d_i^2 \gets \abort$
        \ElsIf{$\Cons{i}{v_j}$ \textbf{ and } $\exists x\in\{0,1,\abort\}$ s.t. $d_j^1=x\;\forall j$}
            \State $d_i^2 \gets x$
        \Else
            \State $d_i^2 \gets d_i^1$
        \EndIf
        \State send $(d_i^2,\ \Pi_2^i)$ to all $j$, where $\Pi_2^i$ is the set of proofs used to justify $d_i^2$ (e.g., $\{v\}\cup\{v_j:\Cons{i}{v_j}\}$)
        \State wait to receive $R_2^i=\{(d_j^2,\Pi_j)\}_{j\in \Lts\setminus\{i\}}$
        \vspace{.25em}
        \Statex \textbf{Round 3}
        \If{$d_i^2\neq \abort$ \textbf{ and } $\exists j,\ v',v''\in\Pi_j$ s.t. $d_2^1=\abort$, $\Cons{i}{v'}$, $\Cons{i}{v''}$, and $\contrad{v'}{v''}$}
            \State $d_i^3 \gets \abort$
        \ElsIf{$d_i^2\neq \abort$ \textbf{ and } $\exists j\;[\,d_j^1\in\{0,1\},\ d_j^1\neq d_i^1,\ \Cons{i}{v_j}\,]$}
            \State $d_i^3 \gets \abort$
        \Else
            \State $d_i^3 \gets d_i^2$
        \EndIf
        \State \textbf{return} $d_i^3$
    \EndProcedure
\end{algorithmic}
\end{algorithm}

\section{Simulations and Design of Experiments}


We use the Aliro Quantum Network Simulator (AQNSim) \cite{aliro-simulator} for the discrete event simulation of the EPRQDBA protocol. We used the NCSA Delta CPU computing resource to conduct a sweep over simulation parameters. A single instance of the simulation only takes about 5 GB of memory, but the simulation runtime varies greatly with different parameter configurations. Due to the stochastic nature of the protocol, all configurations tested in our experiments produce ensemble outcomes to ensure their reliability. We explore the effect of noise on three system setups: a simulation of a logical quantum network, a superconducting network, and a photonic network. 

\subsection{Implementation}

%



The AQNSim Python package provides a discrete event simulation framework for quantum networks. In this framework, the quantum physics of each network device and hardware component is modeled explicitly in time as the EPRQDBA protocol is performed. AQNSim explicitly models the timing of events, allowing us to emulate the performance of the EPRQDBA in a quantum network with latencies and loss. Furthermore, we accurately model the various noise models that arise in quantum computing and networking settings.

AQNSim can be compared to discrete event simulators for quantum networks, such as Netsquid \cite{Coopmans2021_netsquid} or  SeQUeNCe \cite{Wu2021_sequence}.
In general, discrete event simulation of quantum systems extends standard quantum circuit simulators with an explicit timeline on which events occur \cite{bel2024quantum_network_simulators}. In many cases, the timing of events is stochastic or not easily predicted, which makes it challenging to model the uncertainty of real-world quantum   networking protocols. Discrete event simulation provides a means by which to model the stochasticity and uncertainty of real-world hardware.
AQNSim is good fit for this work because it can emulate multiagent logic and behaviors in the EPRQDBA protocol while accurately modeling the low-level physical details of entanglement distribution and measurement \cite{chalupnik2025_realistic_bbm92_sim}.

\subsection{Noise models}\label{section:noise_models}

Noise in quantum systems is modeled using completely-positive trace-preserving (CPTP) maps, $\mathcal{N} : D(\mathcal{H}^{in}) \to D(\mathcal{H}^{out})$. Using the operator-sum representation \cite{nielsen2000quantum}, a CPTP map can be described as
\begin{equation}\label{eq:general_channel_def}
    \mathcal{N}(\rho) = \sum_{K \in \mathcal{K}} K \rho K^\dagger,
\end{equation}
$K \in \mathcal{K} \equiv \{K_0,K_1,\dots\}$, and $\sum_{K \in \mathcal{K}} K^\dagger K = \mathbb{I}$. The operators $K\in \mathcal{K}$ are often referred to as Kraus operators.

Pauli channels are a family of quantum noise model in which
\begin{equation}
    \mathcal{P}(\rho) = p_0 \rho + p_x X\rho X + p_y Y\rho Y + p_z Z \rho Z
\end{equation}
where $1= p_0 + p_x + p_y + p_z$. Special cases of Pauli channels include common qubit channels, such as bitflip ($p_y=p_z =0$), phaseflip ($p_x=P_z=0$), dephasing ($p_x=p_y=0$), and depolarizing ($p_x = p_y = p_z$) channels. Although Pauli channels are simple and versatile, they do not fully describe the noisy processes in described by the operator-sum representation in Eq.~\eqref{eq:general_channel_def}.

In superconducting qubit setups, amplitude damping noise and dephasing noise are important sources of error \cite{krantz2019engineering_sc_qubits}. Amplitude damping models the relaxation of the transmon qubit from the excited state $\ket{1}$ to the ground state $\ket{0}$ as the CPTP map
\begin{equation}
    \mathcal{A}_{\gamma_0}(\rho) = K_0 \rho K_0^\dagger + K_1 \rho K_1^\dagger
\end{equation}
where the amplitude damping Kraus operators are
\begin{align}
    K_0 &= \begin{pmatrix}
        1 & 0 \\ 0 & \sqrt{1 - \gamma_1}
    \end{pmatrix} & K_1&= \begin{pmatrix}
        0 & \sqrt{\gamma_1} \\ 0 & 0
    \end{pmatrix}.
\end{align}
The dephasing channel can be modeled as a general Pauli channel $\mathcal{P}(\rho) = p_0\rho + p_z Z\rho Z$, however, it is more conveniently represented by its Kraus operators
\begin{align}
    K_0 &= \begin{pmatrix}
        1 & 0 \\ 0 & \sqrt{1 - \gamma_1}
    \end{pmatrix} & K_1 &= \begin{pmatrix}
        0 & 0 \\ 0 & \sqrt{\gamma_1}
    \end{pmatrix}.
\end{align}
In the setting of a transmon qubit, a composite noise model combines amplitude damping and dephasing as $\mathcal{P}_{\gamma_2}(\mathcal{A}_{\gamma_1}(\rho))$ where the noise parameters, $\gamma_1$ and $\gamma_2$, are time-dependent. In practice the relaxation rate and dephasing rate are characterized by time constants $T_1$ and $T_2$ respectively where, at time $t$ seconds, the amplitude damping parameter is  $\gamma_1 = 1 - e^{-t / T_1}$ while the dephasing noise parameter is  $\gamma_2 = 1 - e^{-t (2T_1 - T_2) / 2 T_1 T_2}$. In the dephasing case, $\gamma_2$ must account for the implicit dephasing introduced via amplitude damping. As a consequence, this model requires that $T_1 \geq T_2 / 2$.

Loss is a major source of noise for optically encoded qubits \cite{wehner2018quantum,cacciapuoti2019quantum}.
In this process a qubit is lost to the environment.
When a single qubit in an entangled pair is lost, its entangled qubit is projected into the maximally mixed state.
Loss can come in two forms, heralded and unheralded. In the heralded case, the loss event is detectable, allowing the event to be disregarded. This means that heralded loss can decrease the rate at which EPRQDBA can be performed, but will not affect the performance of the protocol. In the unheralded case, the loss event is not detectable, meaning that processing continues with the error. This error can affect either the entanglement verification or byzantine agreement protocol. 


\subsection{Design of experiments}
%

First, the EPRQDBA protocol was reimplemented using pairs of qubits for $N=3$ players. This was then generalized to any $N$ number of players, with the centralized Distributor node sending $M$ $(N-1)$-tuples of correlated qubits to each Lieutenant as described above. In our implemented protocol, traitors do not coordinate attacks, but instead make random decisions and communicate them accordingly to the other nodes in the network.

Starting with our generalized logical protocol, we analyze our network in a noiseless setting, then with the addition of Pauli noise, to determine the protocol's probability of error. With a loyal commander, an error is defined as a loyal lieutenant concluding the protocol on a decision other than the sent command, i.e. aborting or the wrong decision. With a disloyal commander, an error is defined as any loyal lieutenant deciding on any decision \textit{other} than aborting the protocol. We perform 10 runs of each simulation, averaging the number of errors that occur in 30 shots of the protocol in each run, resulting in a total of 300 shots averaged at each point.

The protocol is then adapted into superconducting and photonic variations, which are similarly simulated to analyze their variability under noise.

\newcolumntype{S}{>{\raggedright\arraybackslash}p{0.1\linewidth}} 
\newcolumntype{M}{>{\raggedright\arraybackslash}p{0.14\linewidth}}
\newcolumntype{L}{>{\centering\arraybackslash}p{0.6\linewidth}}    

\begin{table}[!htp]
\centering
\caption{Simulation parameters.}
\begin{tabular}{|S|L|M|}
 \hline
 \textbf{Name} & \textbf{Description} & \textbf{Values} \\
 \hline
 $N$  & Number of Players  & 3--11  \\
 \hline
 $T$  & Number of Traitors  & 0--10  \\
 \hline
 $M$  & Number of Qubit Tuples  & 16--160  \\
 \hline
 $P_p$  & Pauli Noise Probability ($p_0, p_X, p_Y, p_Z$)  & {\small $p_0$: 0.975}  \\
 \hline
  $\alpha$  & Optical Attenuation  & {0.02 dB/km}  \\
 \hline
   $L$  & Optical Channel Length  & {10 cm-100 km}  \\
 \hline
    $T_1$  & Superconducting $T_1$ Time  & {0.05ms}  \\
 \hline
     $t$  & Superconducting Transmission Time  & {0.05ns-0.05ms}  \\
 \hline
 $n_r$  & Number of simulation runs  & 10  \\
 \hline
 $n_s$  & Number of Shots in each run  & 30  \\
 \hline
\end{tabular}
\end{table}



\section{Results}


\subsection{Logical Qubits}

\subsubsection{Ideal Case}

In the absence of noise, we analyze the amount of sent qubit-tuples $M$ necessary to sufficiently reduce the probability of error. This gives a minimum number of qubits necessary for robust experimental demonstration of EPRQDBA under varying network traitor densities. In all cases, perfect success is asymptotically approached, but as more traitors are introduced to the network, a larger $M$ is required to reach this performance.

\begin{figure}[H]
    \centering
    \includegraphics[width=0.9\linewidth]{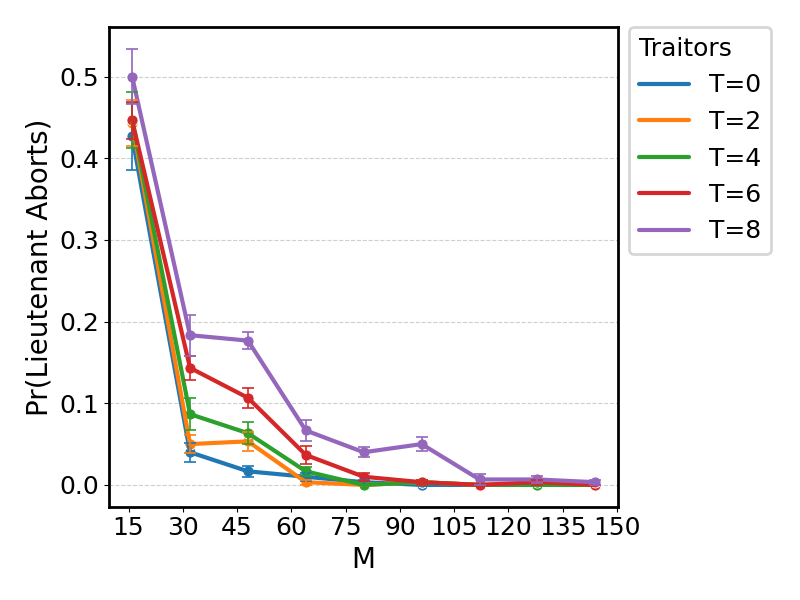} 
    \caption{Noiseless abort probability with loyal commander in varied traitor density networks of size $N=11$.}
    \label{fig:traitors}
\end{figure}


\subsubsection{Pauli Noise}\label{section:pauli_noise}
Pauli noise has four parameters, $p_0$, the probability of no Pauli flip occurring, and $p_x, p_y, p_z,$ the probability of a Pauli flip occurring in the $X$, $Y$, and $Z$ axes respectively. We analyze the protocol's probability of error across different Pauli noise probability distributions to determine the protocol's sensitivity to Pauli noise in the $X$, $Y$, and $Z$ directions with a fixed $p_0$ across different network sizes. The addition of Pauli noise is applied directly before Lieutenants measure their received qubits. Thus, Pauli errors occur during quantum measurement, and are then propagated through classical communication, eventually resulting in possible protocol errors. The following ternary plots fix a value of $p_0=0.975$, then distribute the total probability of error, $1-p_0=0.025$ between the $X$, $Y$, and $Z$ axes, measuring the probability of lieutenant error at each point. Our analysis is conducted over 105 equally spaced points across the 2D phase space of Pauli noise with fixed $p_0$ and $M=112$.

\begin{figure}[H]
    \centering
    \includegraphics[width=0.7\linewidth]{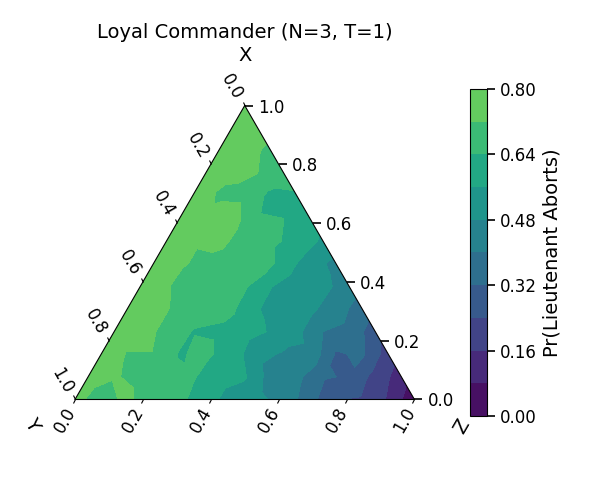}
    \caption{Pauli noise in a logical qubit network of $N=3$ nodes with $T=1$ traitors and a loyal commander.}
    \label{fig:pauli-n3}
\end{figure}

\begin{figure}[H]
    \centering
    \includegraphics[width=0.7\linewidth]{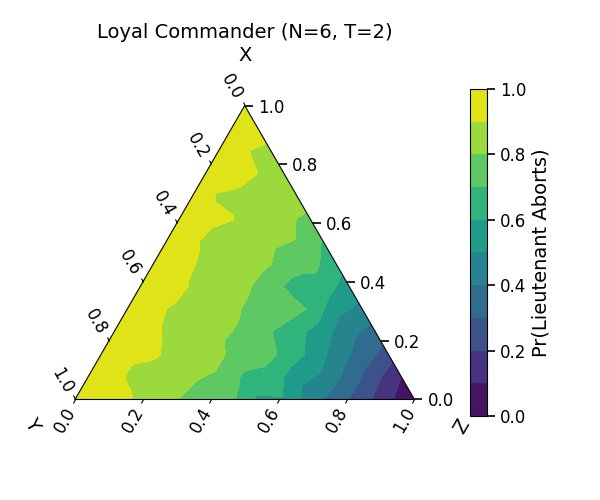}
    \caption{Pauli noise in a logical qubit network of $N=6$ nodes with $T=2$ traitors and a loyal commander.}
    \label{fig:pauli-n6} 
\end{figure}

\begin{figure}[H]
    \centering
    \includegraphics[width=0.7\linewidth]{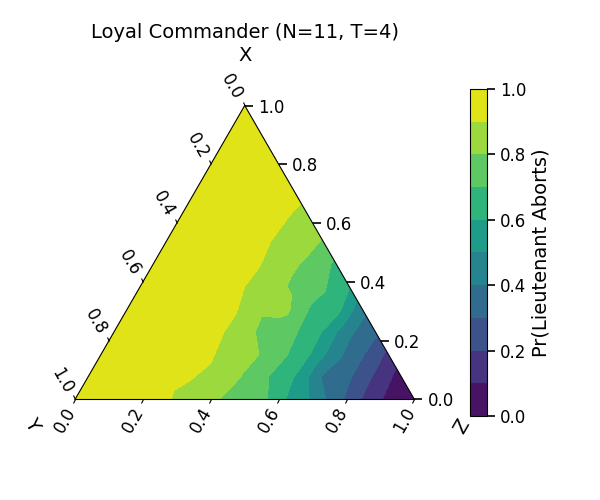}
    \caption{Pauli noise in a logical qubit network of $N=11$ nodes with $T=4$ traitors and a loyal commander.}
    \label{fig:pauli-n11}
\end{figure}

As network size increases, the effects of Pauli noise become more severe in causing protocol errors. Pauli $Z$ noise, resulting in only qubit state phase changes, has no effect on the protocol, while $X$ and $Y$ noise, which both result in logical state flips, drastically reduce protocol success. In larger networks, this Pauli noise is propagated further through more pairwise lieutenant communications during the protocol, amplifying its increasing effects on the probability of error.

Interestingly, with a traitorous commander, the protocol's success (all loyal Lieutenants aborting the protocol) increases with larger amounts of effective pauli noise. Protocol success is highest with entirely $Z$ (dephasing noise), where the protocol works near-noiselessly. However, the introduction of slight Pauli noise in the $X$ and $Y$ directions significantly decreases the probability of network success, before increasing near maximum effective noise along the $XY$ axis. 

\begin{figure}[H]
    \centering
    \includegraphics[width=0.7\linewidth]{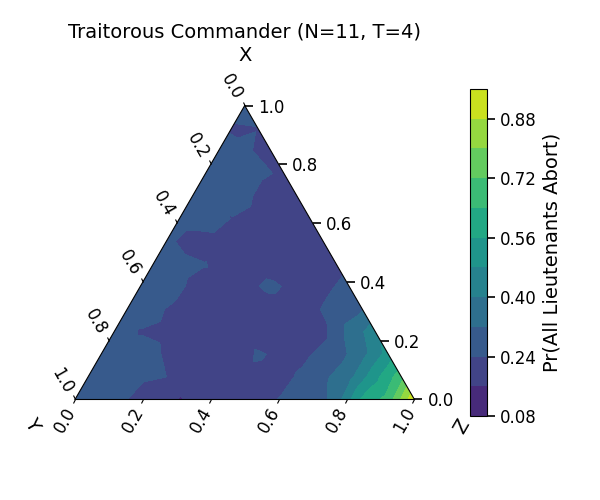}
    \caption{Pauli noise in a network of $N=11$ nodes with $T=4$ traitors and a traitorous commander.}
    \label{fig:traitorous_commander}
\end{figure}

\subsection{Superconducting qubits}

We model a modular and distributed quantum processor in which deterministic quantum state transfer is used to prepare entangled states between remote superconducting qubit processors \cite{Kurpiers2018_microwave_deterministic_state_transfer,campagne2018_sq_determinstic_remote_entanglement,Zhong2019_sc_bell_violation,magnard2020_microwave_sqc_link,Zhong2021_sq_deterministic_entanglement,burkhart2021_sqc_error_detected_state_transfer,Niu2023_low_loss_interconnects_sc,grebel2024_sqc_bidir_state_transfer}. Recent advances in quantum state transfer between remote superconducting qubit processors have shown to produce entangled states $\ket{\Phi^-} = \frac{1}{\sqrt{2}}(\ket{00} - \ket{11})$ with a fidelity around 99\% and a preparation time of about 25 nanoseconds \cite{Niu2023_low_loss_interconnects_sc}. Furthermore, entanglement purification can be applied in superconducting qubit systems to improve the state preparation fidelity, or maintain high fidelity over longer durations \cite{yan2022_sc_qubit_remote_purification}.

As discussed in Section~\ref{section:noise_models}, the main source of error for superconducting qubit setups is decoherence over time. The decoherence process is modeled using a $T_1$ (amplitude damping) and $T_2$ (dephasing) relaxation rate.
As shown in Section~\ref{section:pauli_noise}, phase errors do not affect the EPRQBA protocol because the lieutenants immediately measure incoming qubits from the distributor, and only anticorrelation between entangled pairs is used in the protocol rather than any phase information. Therefore the EPRQDBA protocol is invariant under $T_2$ relaxation (dephasing noise). On the other hand, $T_1$ relaxation (amplitude damping noise) does have an impact. We look at the commonly studied $T_1$ noise as a source of amplitude-related noise in the superconducting case. We study the effect of noise by increasing the ratio of the superconducting wire length with respect to the $T_1$ noise. We first explore the probability that a loyal node aborts the protocol when the commander is loyal. We also explore the effect of noise on detecting a traitorous commander.

\begin{figure}[h]
    \centering
    \includegraphics[width=0.9\linewidth]{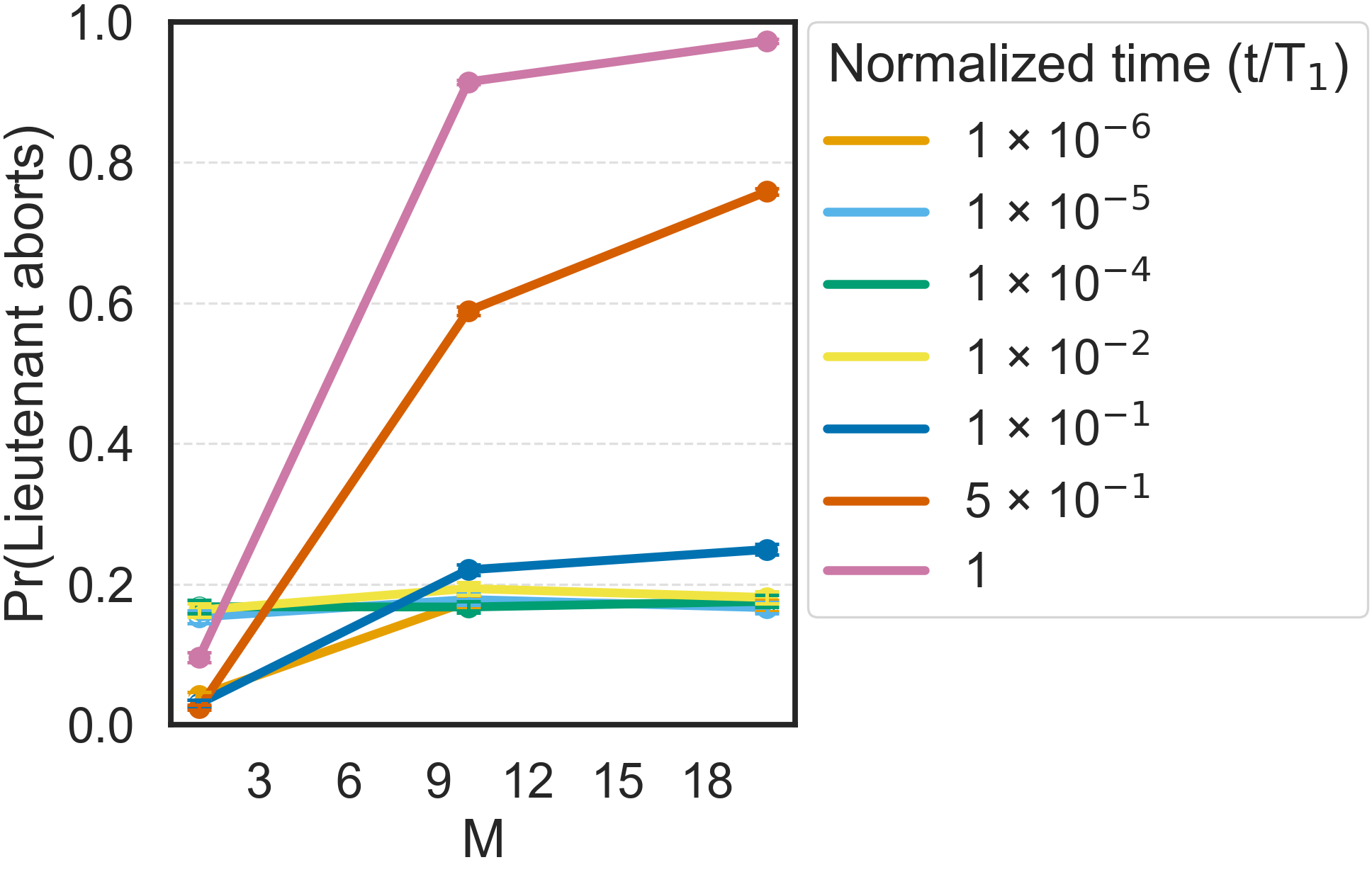}
    \caption{$T_1$ noise in a superconducting qubit network with 5 nodes and a loyal commander, averaged over $N \in \{3,5,10,15\}$.}
    \label{fig:superconducting-M}
\end{figure}

\begin{figure}[h]
    \centering
    \includegraphics[width=0.9\linewidth]{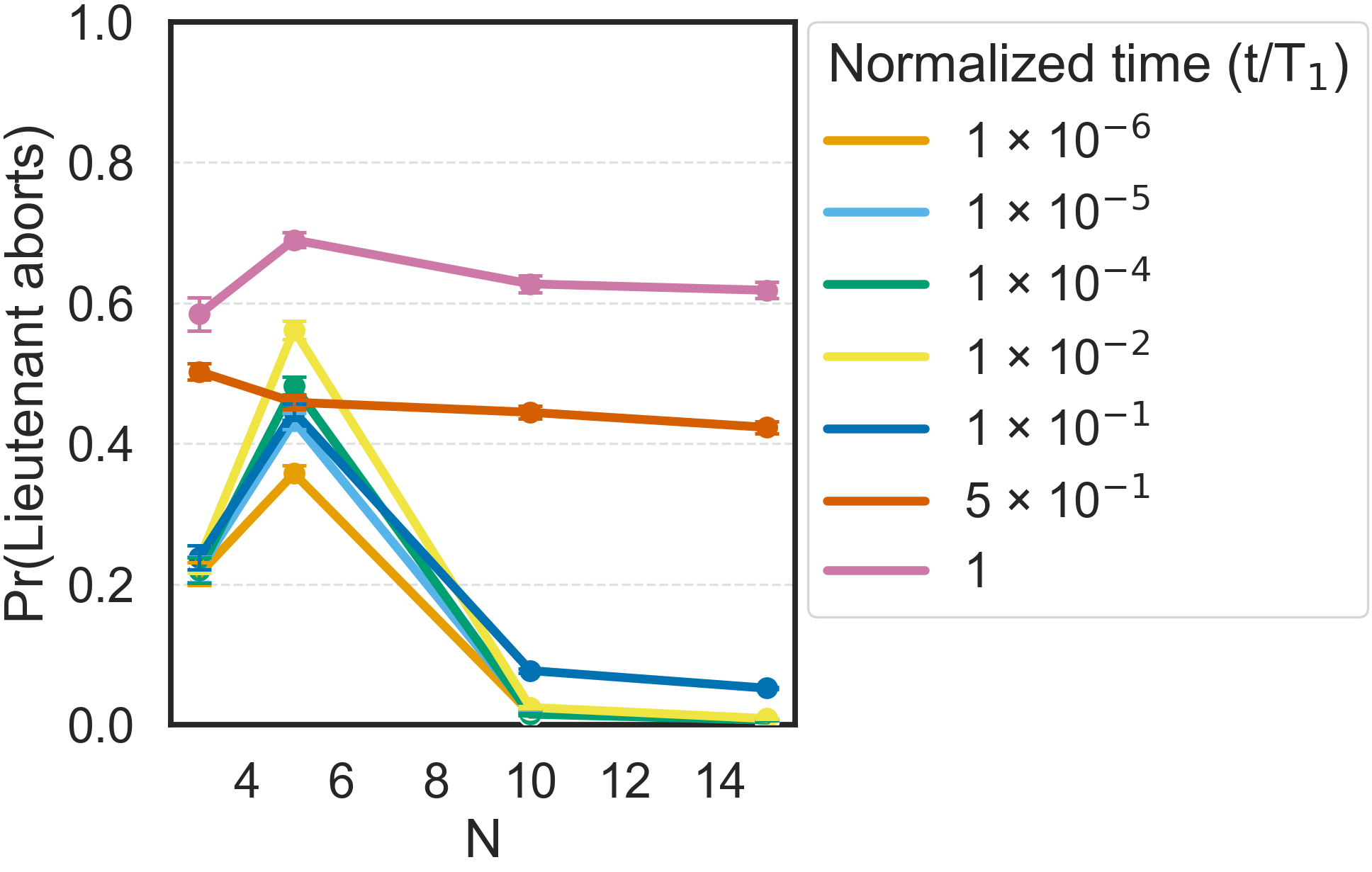}
    \caption{$T_1$ noise in a superconducting qubit network with no traitors and a loyal commander, averaged over $M \in \{1,10,20\}$. The number of nodes $N$ is varied.} 
    \label{fig:superconducting-N}
\end{figure}

\begin{figure}[h]
    \centering
    \includegraphics[width=0.9\linewidth]{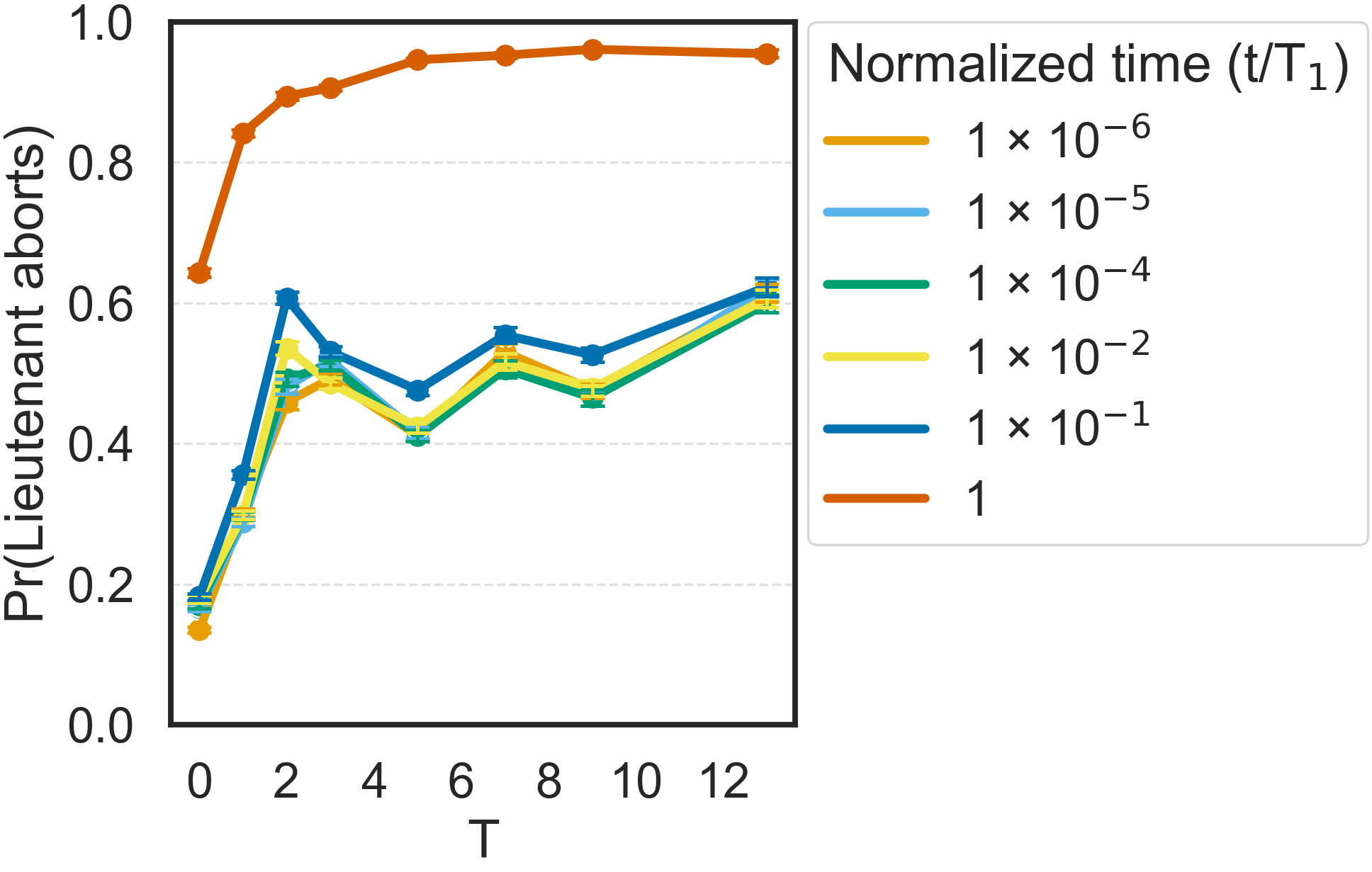}
    \caption{$T_1$ noise in a superconducting qubit network with 5 nodes and a loyal commander. The multiplicity of entanglement distribution $M$ is varied.}
    \label{fig:superconducting-traitor-M}
\end{figure}

\begin{figure}[h]
    \centering
    \includegraphics[width=0.9\linewidth]{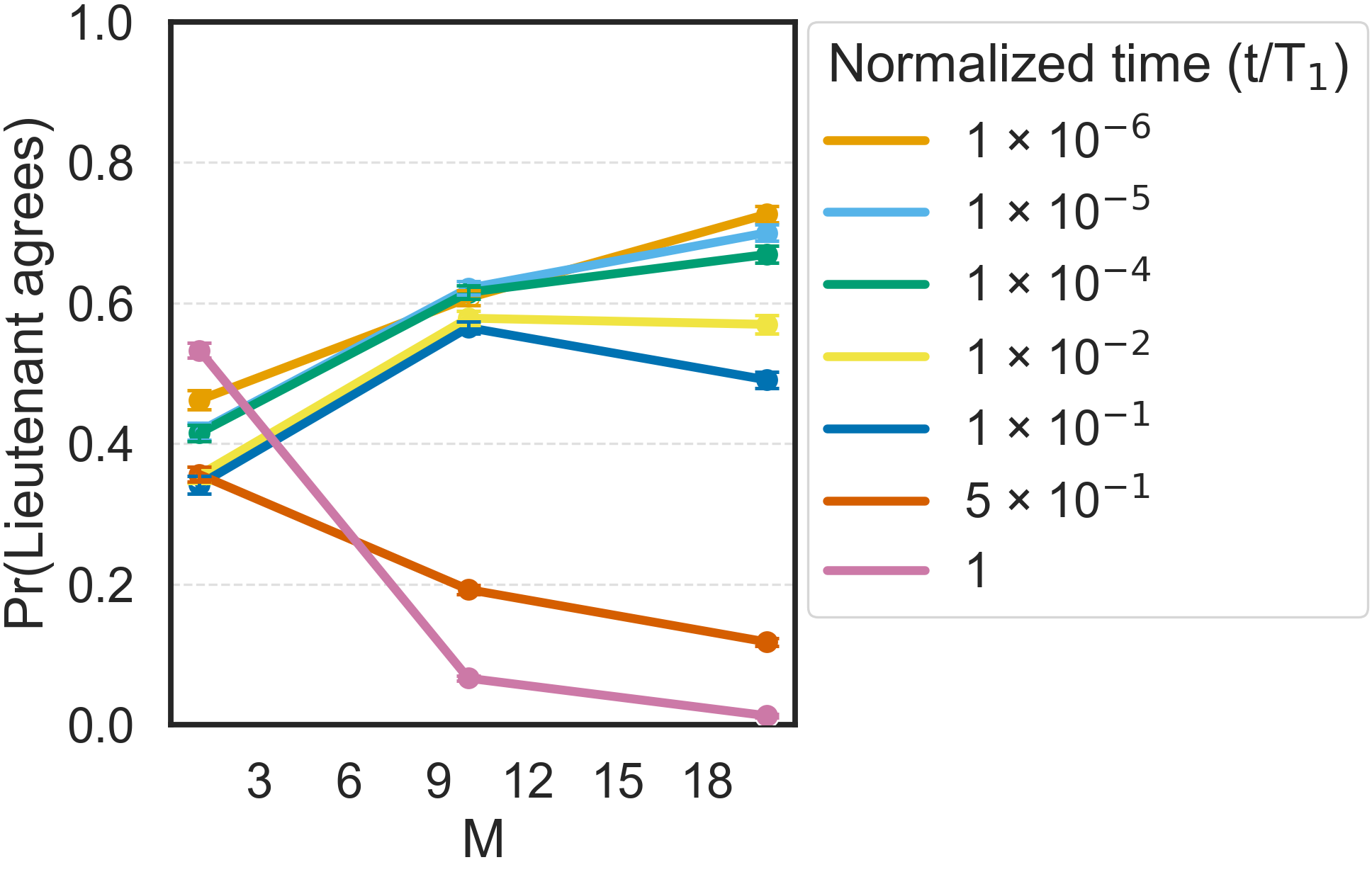}
    \caption{$T_1$ noise in a network with 5 nodes and a traitorous commander, averaged over $M \in \{1,10,20\}$. The number of traitors $T$ is varied.}
    \label{fig:superconducting-traitor-M}
\end{figure}

\subsection{Photonic qubits}

Photonic setups are a practical approach for performing Byzantine agreement protocols because photonic qubit states are well-suited for long-distance transmission of quantum information and photonic qubits are easier to prepare, manipulate, and detect than superconducting qubits or other matter based qubit setups.

\begin{figure}[t]  
    \centering
    \includegraphics[width=\linewidth]{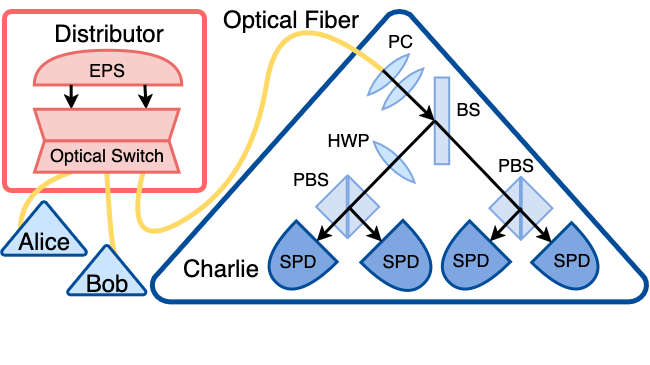}
    \caption{A schematic of a photonic qubit network in which polarization qubits are used to perform EPRQDBA among three parties Alice, Bob, and Charlie. In each round, the distributor produces an entangled pair from the  entangled photon source (EPS). The entangled photons are  routed through an optical switch to the appropriate player, traveling along a fiber optic cable from the distributor to each player's measurement device. The measurement device has a polarization controller (PC) to correct polarization drift and  a beam splitter (BS) to passively select between basis measurements at random. A half wave plate (HWP) is used to setup $X$ and $Z$ basis measurements while a polarization beam splitter (PBS) followed by single-photon detectors (SPD) are used to perform a measurement.   }
    \label{fig:optical-network-schematic}
\end{figure}

The main components needed for a successful optical qubit EPR-based QDBA protocol are a photonic entanglement source, an optical switch, fiber optic cables (or alternatively, a free-space communication system), and single-photon detectors (see Fig.~\ref{fig:optical-network-schematic}). Each of these components are explicitly modeled in the AQNSim discrete event simulation framework, which allows for accurate simulation of photonic systems.

We consider polarization encoded qubits, in which quantum information is encoded into a photon's polarization. In this encoding, a qubit is represented by a single-photon excitation in the state $\ket{\psi}=a\ket{H} + b\ket{V}$ where $|a|^2 + |b|^2 = 1$. The computational basis or $z$-basis corresponds the photon having a horizontal or vertical polarization, $\ket{H}$ or $\ket{V}$, where the relative phase between the two corresponds to the photon having an elliptical polarization. In practice, entangled polarization qubits can be prepared via spontaneous parametric down conversion where optical components process the entangled qubits. Projective qubit measurements can be performed locally by each party using polarizing beam splitter detector setups. We modeled the unheralded loss case, in which any lost qubit was assigned a bit value of 0, as if it had been measured in the $\ket{H}$ state. Although we focus on polarization-encoded qubits, alternative options encodings, such as frequency-bin or presence-absence, could be used for EPRQDBA.

To calibrate each party's measurement bases and perform entanglement verification, the measurement basis must be rotated. For polarization qubit encodings, the basis selection can either be actively controlled by rotating wave plates, or passively controlled using a beam splitter to select between two measurement bases.

\begin{figure}[h]
    \centering
    \includegraphics[width=0.9\linewidth]{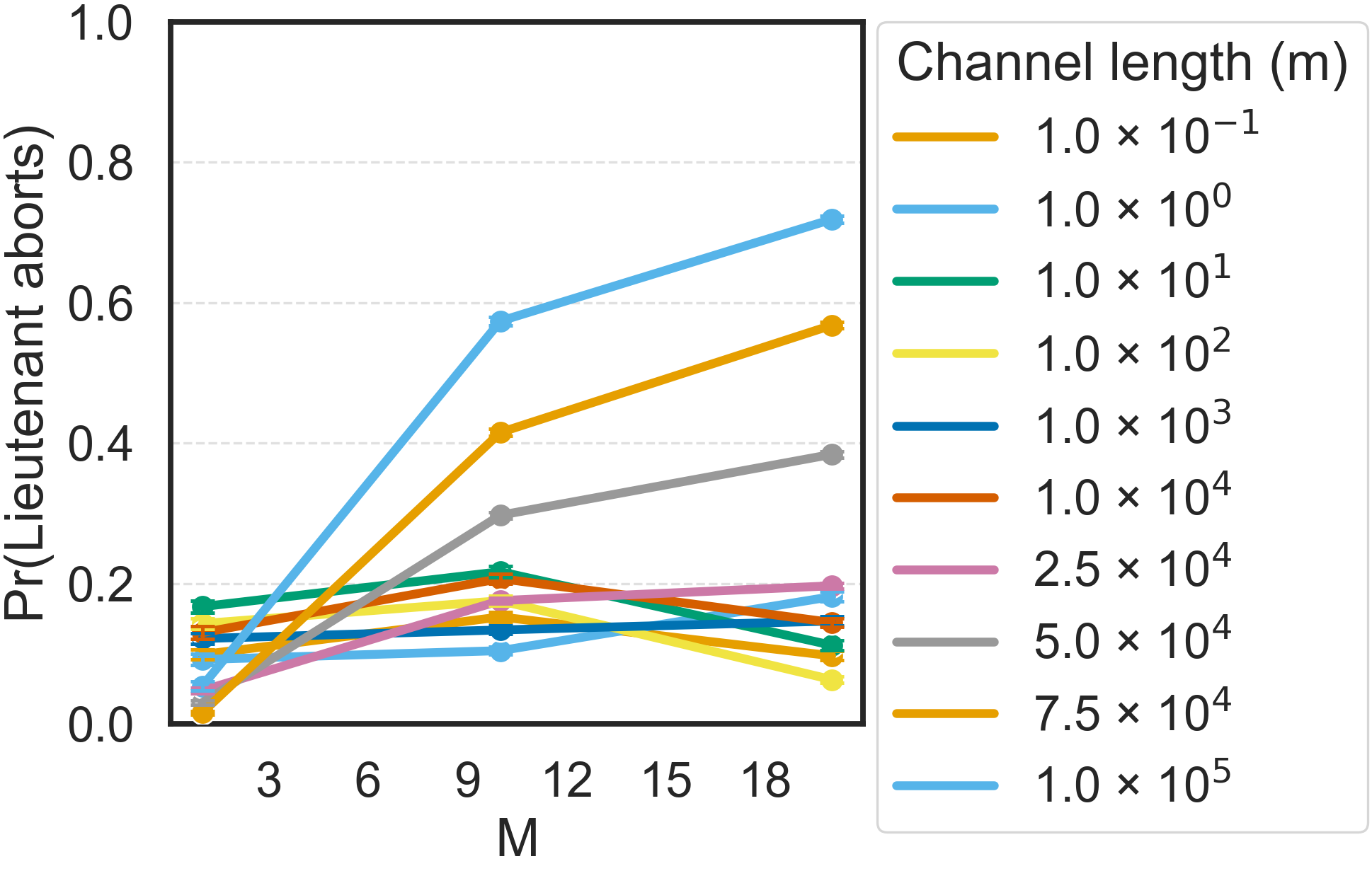}
    \caption{Attenutation in a photonic qubit network with no traitors and a loyal commander, averaged over $N \in \{3,5,10,15\}$.}
    \label{fig:optical-M}
\end{figure}

\begin{figure}[h]
    \centering
    \includegraphics[width=0.9\linewidth]{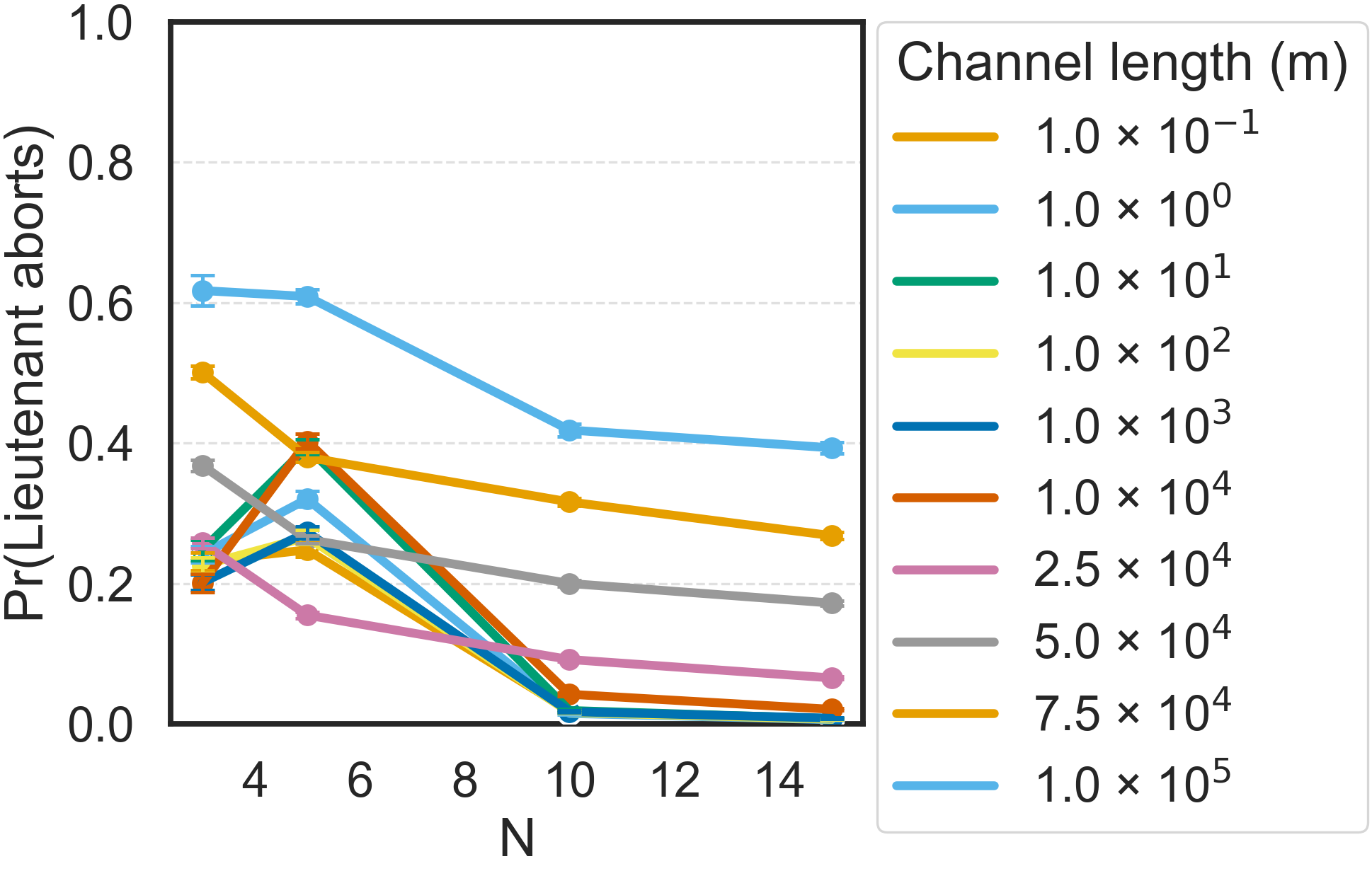}
    \caption{Attenutation in a photonic qubit network with no traitors and a loyal commander, averaged over $M \in \{1,10,20\}$.}
    \label{fig:optical-N}
\end{figure}

\begin{figure}[h]
    \centering
    \includegraphics[width=0.9\linewidth]{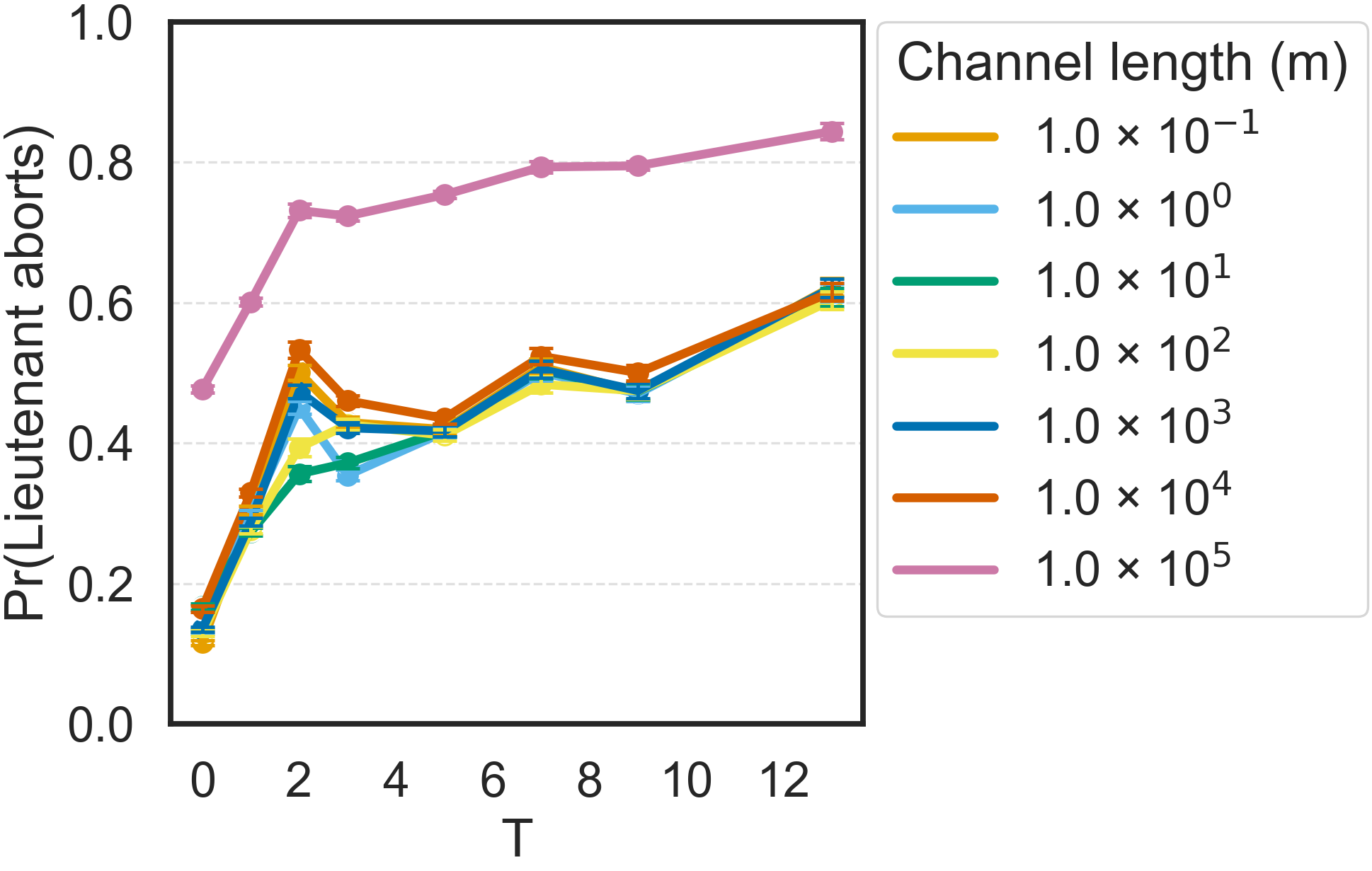}
   \caption{Attenuation in a photonic qubit network with 5 nodes and a loyal commander, averaged over $M \in \{1,10,20\}$. The number of traitors $T$ is varied.}
    \label{fig:optical-T}
\end{figure}

\section{Discussion and Significance}
Our results quantitatively analyze the parameter space of the EPRQDBA protocol. First simulating without noise, we characterize the minimum $M$ necessary to achieve consensus with high probability at a given traitor density. Fixing $N=11$, for example, we determined that a minimum $M=112$ ensures complete success for any $T$. Successful agreement at any $T$ is the fundamental benefit of QDBA over classical protocols, and ensuring a sufficiently large $M$ is used for a given network guarantees quantum advantage in BA. This is illustrated in the 3-player case of the system, the most basic configuration of BA that is classically unsolvable. We see in Figure \ref{fig:pauli-n3} that without harmful noise (i.e. noise probability is mostly relegated to $p_z$), there is a 0\% probability of error, demonstrating the loyal Lieutenant always agrees with the loyal Commander. As the amount of harmful noise increases, so does the probability of ta Lieutenant aborting. We observe that noise never causes traitor success in which a Lieutenant  decides on an incorrect command instead of aborting. This is the most important aspect of QDBA: if a node agrees on a command, it is guaranteed to be the correct command, maintaining the validity in the network.
In the case of protocol abortions due to noise, repeating the protocol until consensus is achieved will result in the same end result, just requiring more time and resources. This is relevant for designing robust decentralized systems, where reliability is the most important consideration. 

Next we generalize to the $N$-player case, by scaling our simulations to larger networks with $N=6$ (Fig.~\ref{fig:pauli-n6}) and $N=11$ (Fig.~\ref{fig:pauli-n11}) where  the traitor densities are at least one third. Our simulation results highlight the propagation of Pauli noise as $N$ increases. Although the probability of lieutenant abortion blows up as the Pauli noise mixture approaches the $XY$ axis, the density of complete protocol success close to pure $Z$-noise increases with $N$. This suggests that the EPRQDBA protocol's ability to establish BA does not degrade as $N$ increases and the ratio of traitors remains constant. In fact, our simulations demonstrate that  increasing the number of parties can cause more extreme loss when $X$ and $Y$ noise is high, while a larger region of high success probability is found near high $Z$ noise.

With a traitorous commander (Fig.~\ref{fig:traitorous_commander}), we see our minimum success rate falls in the middle of our Pauli noise distribution, as opposed to the $XY$ plane as found in loyal commander figures. This is due to significant Pauli noise causing a larger number of Lieutenants to abort from receiving contradictory information, which is the correct decision in the case of a disloyal commander. These abortions result in a lower probability of error, and therefore, a higher success rate.

In the photonic and superconducting simulations, we see very similar results. This is because phase noise does not interfere with the protocol, so only amplitude damping and loss affect success probabilities. The impact of loss is exponential, so we see suddenly degrading performance metrics. However, for experimentally expected degrees of noise (e.g., $T_1 > t_{\text{transmission}}$ and $\text{Channel Length} < 100 \text{ km}$) we do not reach the regime of sudden performance drop-off.

Our work demonstrates that rigorous simulations of protocols on quantum networks are useful even for a small number of players. First, simulation provides an avenue for validating both physical intuitions and formal expected results under variation of parameters. Second, and most importantly, our findings show the presence of counterintuitive results resulting from the synergistic interaction of multiple parameters, as well as from the peculiarities involved in modeling and building true distributed systems. Finally, our work demonstrates how quantum network simulators decouple the execution of in-lab experiments from protocol design, providing a natural division of labor while preserving essential assumptions about quantum hardware during co-design phases. By modeling properties of noise tied to hardware properties, our models have a better chance at being realistic and translatable to quantum network testbeds. We expect the latter to accelerate research at the intersection of quantum information theory and quantum networks. Clearly, realistic quantum networks simulators become essential to accomplish this.

Distributed quantum computing involves two main classes of research from the perspective of protocol execution. One line of research explores individual, simple protocols to assert whether the necessary guarantees hold and why. This involves, for instance, understanding the role of quantum resources under less than ideal conditions in line with experiments. The second line of research, much less explored due to how expensive quantum network experiments become as complexity increases, is the analysis of how multiple quantum protocols compose resulting in larger ones useful for actual applications. Larger, in this context, also involves scaling up the number of nodes in the network both classical and quantum, as well as the number of simultaneously concurrent protocols in it. Sufficiently realistic simulations provide a unique window into these regimes, which in turn becomes a relevant challenge for the discrete event simulation and parallel computing communities regarding enabling cyberinfrastructure.

Finally, in the process of studying the protocol described in \cite{EPRQDBA}, we observed that relying on a common coin protocol to establish a shared random secret is key to substantiating both its security and its ability to solve for $n = 3$. For that to be the case, distributed BA must be guaranteed to remain correct were classical shared secrets available using classical common coin protocols. Feldman and Micali showed precisely that BA and DBA are reducible to such a protocol \cite{feldman1988optimal}, thus providing a stronger basis to start from than the original literature cited by the article whose results we reproduce. The significance of the latter resides in the ability to bring stochastic game theoretic considerations directly, while observing that entanglement in the form of EPR pairs results in solvability of the minimal case stated above.

Even more significantly, the protocol described above has peculiarly few security prerequisites. For instance, DBA does not require trusted parties or preprocessing, achieving guarantees for $t < n/3$ faults when synchronous is present, and $t < n/4$ faults for the asynchronous case. To achieve this, BA protocols must ensure that all good processors have the ability to coordinate effectively, which involves \emph{consistency} (i.e., the ability to agree on the same value repeatedly), and \emph{meanignfulness} (i.e., the ability to remain consistent from the start) in the presence of either channel errors or malicious attacks. Feldman and Micali point out sharply that a large share of DBA protocols fail to reach consensus when these conditions are not met (``DBA from scratch'' ). The EPRQDBA described here constructs an analogous common random coin through the graded broadcast and graded verifiable secret sharing protocols, which play a similar role as the command and proof vectors in our case.

In general, the results obtained here in coordination with the observations raised by Feldman and Micali suggest that once a common coin is available, aspects such as the availability of private channels and authentication appear secondary to the ability of ensuring either trustable coordination or defecting in the light of doubt. Additionally, we believe the solvability of $n=3$ is due to the quantum task of shared secret generation using a public/faulty entanglement source. Moreover, it is left to further research how classical and quantum DBA protocols may sit at different points between purely algorithmic hardness and physical security to ensure a trusted outcome.

\section{Conclusions}

In this work, we sweeped the parameter space of the EPRQDBA protocol to serve experimentalists with a concrete reference of theoretical results. We find that there are severe performance degradations for excessive quantities of noise which are unlikely to appear in most experimental setups. Overall, the EPRQDBA protocol is quite robust. With a loyal commander, each loyal lieutenant is likely to make the optimal decision for limited quantum resources: $M, N$. With a traitorous commander, increasing $M$ is paramount to detecting a Byzantine order.

The most salient fact uncovered by our research is the unexpected role that noise plays in the presence of defectors. Noise only helped the protocol when the commander is traitorous, since it increases the probability of collective defection, thus safeguarding all involved actors. This behavior appears to be substantially different from the common coin protocol described above as well. Incidentally, $N$ appears to take a role as a modulator of the effect of noise, exacerbating its effects either on agreement or defection at the large limit. In summary, the security of the EPRQDBA protocol is robust to the presence of noise.

Future work stemming from this article involves three main directions. First, investigating immediately available improvements to QDBA by determining if optimizations in the original common coin protocol found during our research do indeed apply to the present case. Second, we would like to explore new QDBA protocols with better agreement probability by using quantum resources beyond entanglement, such as the utilization of phases to go beyond immediate defection. We believe the latter would help reduce false positives during detection of foul play, and it would enable parametric sweeps with Pauli $Z$ noise to study its effects on the degradation of agreement. Third, our simulations can be contrasted against experimental setups for superconducting qubits, photonic qubits, or other distributed quantum systems. Finally, whether protocols exist for QDBA that actually perform quantum computations in a distributed manner and with better security or convergence properties remains an open question.

\section*{Code and Data Availability}

All supporting code and data is publicly available on GitHub at \href{https://github.com/DistributedQC/Byzantine-Agreement}{https://github.com/DistributedQC/Byzantine-Agreement}. Access to the AQNSim package is necessary for running simulations.

\section*{Acknowledgments}

This work is partially funded by the NSF QLCI Hybrid Quantum Architectures and Networks (HQAN: NSF \#2016136) center and Aliro Technologies. Our simulations were run using the Aliro Quantum Network Simulation (AQNSim) Python framework. Our research used the Delta advanced computing and data resource which is supported by the National Science Foundation (award OAC 2005572) and the State of Illinois. Delta is a joint effort of the University of Illinois Urbana-Champaign and its National Center for Supercomputing Applications. Delta access was provided through ACCESS, an advanced computing and data resource program supported by the U.S. National Science Foundation (NSF) under the Office of Advanced Cyberinfrastructure awards \#2138259, \#2138286, \#2138307, \#2137603 and \#2138296.

\bibliographystyle{ieeetr}
\bibliography{references}

\end{document}